\documentclass[fleqn,usenatbib]{mnras}

\usepackage{epsfig}
\usepackage{epstopdf}
\usepackage{subfloat}
\usepackage{float}
\usepackage{subfigure}
\usepackage{hyperref}
\usepackage{multirow}
\usepackage{graphicx,times}  
\usepackage{soul}
\usepackage{amssymb,amsmath}
\usepackage{color}
\usepackage{caption}
\usepackage{comment}
\usepackage[T1]{fontenc}

\def\apjl{APJL}
\def\aap{AAP}

\newcommand{\id}{{\rm d}}

\newcommand{\vt}{{\vec \theta}}
\newcommand{\vU}{{\vec{ U}}}

\def\HI{{H~{\sc i} }}

\def\hi{\rm H \scriptscriptstyle I}

\begin{document}

\title{Large-scale turbulence cascade in  the spiral galaxy  NGC~6946}
\author[Meera Nandakumar and Prasun Dutta ]
{Meera Nandakumar$^{1,2}$\thanks{Email: meeranandakumar93@gmail.com}, 
Prasun Dutta$^{2}$\thanks{Email:pdutta.phy@itbhu.ac.in},  
\\$^{1}$ Department of Physics, Indian Institute  of Science, Bangalore, 560012  India. 
\\$^{2}$ Department of Physics, IIT (BHU) Varanasi, 221005  India. 
}
\maketitle 

\begin{abstract}
The generation mechanism of compressible fluid turbulence at kiloparsec scales in the Interstellar Medium (ISM) is a long-lasting puzzle. In this work, we explore the nature of large-scale turbulence in the external spiral galaxy NGC~6946. We use the Visibility Moment Estimator (VME) to measure the \HI column density and line of sight turbulent velocity power spectra combining the new observations of A array configuration of Karl G. Jansky Very Large Array (VLA) with the VLA B, C, D array observations from The \HI Nearby Galaxy Survey (THINGS). The estimated power spectra are obeying a power law with a slope of $-0.96\pm0.05$ in column density and $-1.81\pm0.07$ in line of sight velocity in  length scales ranging from $6$ kpc to $170$ pc. This points towards a forward energy cascade in the plane of the disc with a driving scale at least as large as $6$ kpc. The values of the power law indices indicate a combination of solenoidal and compressive force responsible in driving the measured turbulence. The presence of strong regular magnetic fields from the magnetic spiral arms in the galaxy is possibly contributing to the solenoidal part, while self-gravity or gravitational instability can mostly be the input for the compressive part of the forcing in the driving mechanism. 
	\end{abstract}


\section{Introduction}
Compressible turbulence in the Interstellar Medium (ISM) plays an important role in governing galactic morphology, dynamics and chemistry. Compressibility in the turbulent ISM results in the hierarchical formation of scale-invariant structures through which energy cascades in a forward manner from large scales (i.e.,  few tens of kiloparsec) to smaller scales where star formation is dominant \citep{1951ApJ...114..165V}. Through different statistical estimators like probability density function, structure-function, auto-correlation, power spectrum etc, turbulent generated coherent structures are probed in our Milky Way galaxy at few parsec scales \citep{1983AA...122..282C,1985A&A...146..223C,1993MNRAS.262..327G,2000ApJ...543..227D,2001ApJ...561..264D,2003A&A...411..109M,2010MNRAS.404L..45R,2010ApJ...716..433K,2010A&A...518L.104M,2012AAS...21932005R,2013ApJ...779...36P,2015ApJ...810...33C,2016A&A...593A...4M,2018ApJ...856..136P,2021ApJ...908..186M} and in some neighbours like Large Magellanic Clouds, Small Magellanic Clouds and some dwarf galaxies at few kpc scales 
\citep{1999MNRAS.302..417S,2001ApJ...548..749E,2006MNRAS.372L..33B,2008MNRAS.384L..34D,2009MNRAS.398..887D,2012ApJ...754...29Z,2017AJ....153..163M,2022PASA...39....5P}. The power spectrum of density and velocity fluctuations coming from the scale-invariant structures obeys a power law in the length scale region where the energy cascade happens. The possible driving mechanism behind the generation of compressible turbulence in the ISM can be inferred from the power law indices. 
For instance, \cite{2010ApJ...714.1398C} estimated the velocity statistic of atomic hydrogen emission from Milky Way at high latitudes and interpreted the obtained steep velocity power law at $\sim 100$ pc scales as being due to the shock-dominated turbulence. Following more similar studies using multi-wavelength observations in Milky Way have been conducted and their results along with various simulations suggest that stellar feedback is the possible driver of turbulence at sub parsec scales \citep{1996ApJ...467..280N,2004RvMP...76..125M,2004Ap&SS.289..479D,2006ApJ...653.1266J,2008ApJ...686..363H,2009A&A...504..883B,2010ApJ...713.1376F,2013MNRAS.429.1437R}. However at the larger scales of order that is of kiloparsec, the scenario is different and is still not clearly understood.

Power spectrum estimation 
of atomic hydrogen (\HI) intensity fluctuations of numerous spiral and dwarf galaxies by \cite{2013NewA...19...89D} and \cite{2009MNRAS.398..887D} respectively shows power-law behaviour at scales ranging from $1$ kpc to $10$ kpc. The power law power spectrum obtained in their sample indicate the presence of scale-invariant structures in the ISM that is attributed to the two-dimensional compressible turbulence. Commonly from the \HI intensity distributions, power spectra of column density and line of sight turbulent velocity can be estimated, where the latter has information about the driving mechanism of turbulence. Methods like 
velocity centroids, Velocity Channel Analysis (VCA) and Velocity Coordinate Spectrum (VCS) have been used in literature widely for estimating velocity power spectrum in our Galaxy and dwarf neighbours like SMC and LMC \citep{2000ApJ...537..720L,2001ApJ...555L..33P,Esquivel_2005,2006ApJ...652.1348L}. However, \cite{2015MNRAS.452..803D} showed that these techniques are not efficient in estimating the velocity power spectrum in external spiral galaxies. 
Following, \cite{2016MNRAS.456L.117D} introduced a method which estimates \HI column density and line of sight turbulent velocity power spectrum using moments of visibility. \cite{2020MNRAS.496.1803N} implemented a variant of the former, named Visibility Moment power spectrum Estimator (VME) in external spiral galaxy NGC~5236. They measured large-scale turbulence cascades starting from a driving scale of $6$ kpc to $300$ pc and conclude that it is possibly driven by gravitational instability or self-gravity. 
\cite{2005ApJ...630..238D} shown through autocorrelation length estimation, the presence of turbulence cascade with input energy injected at a scale of $6$ kpc in dwarf irregular galaxy Holmberg II. Through theoretical modelling, \cite{2016MNRAS.458.1671K} investigated the correlation between star formation rate, gas dispersion and gas fraction in galaxy models where gravity-driven and feedback-driven turbulence is separately considered. Compared with the observational results on a collection of galaxies, they show that predicted correlations from the gravity-driven model is more agreeing with the observation. In later advanced modelling of disc galaxies, \cite{2018MNRAS.477.2716K} shows that for gas-rich high redshift galaxies, gravity-driven turbulence is required to provide the observed gas dispersion while in low redshift galaxies, both feedback and gravity give similar dispersion. In a hydrodynamic simulation of LMC-sized galaxies, \cite{2010MNRAS.409.1088B} shows that large-scale turbulence is regulated by gravitational instabilities.  Recently \cite{2023arXiv230113221F} study the evolution of gravity-driven isothermal turbulent cascade in disc galaxies and show that velocity fluctuations obey Burger's scaling. The slope of the velocity power spectrum in their analysis agrees with the estimations of \cite{2020MNRAS.496.1803N} in NGC~5236 within 2 $\sigma$ uncertainties. %More details on the comparison with the new results from this work are discussed in section~\ref{discu}.

 In this paper, we use the VME estimator to measure \HI column density and line of sight turbulent velocity power spectrum of spiral galaxy NGC~6946. We also compare the earlier results with NGC~5236 and try to make connections between them.
NGC~6946 is a bright late-type spiral galaxy. The atomic hydrogen distribution and rotation curve of this galaxy was initially measured by \citep{1968ApJ...154..845G} using \HI observations from a $300$ foot radio telescope in National Radio Astronomy Observatory. Later \cite{1986ApJ...308..600T} with the \HI observation using Karl G. Jansky Very Large Array (VLA) telescope in the D configuration, detected  \HI emission in the disk that extends upto a radius of $ \sim 30$ kpc. \cite{1996Natur.379...47B} detected two symmetric bright magnetic spiral arms between the optical spiral arms, in NGC~6946 using radio polarization observation. NGC~6946 has well-defined spiral structures and is found to have \HI structures like holes and high-velocity gas \citep{1993A&A...273L..31K,2008A&A...490..555B}. Recently \cite{2022arXiv220509069K} showed that the observed \HI gas rotation curve of NGC~6946 is fitted better with a model where the dynamical effect caused by the magnetic fields in the magnetic arms is also taken into account.
 \cite{2009MNRAS.398..887D} measured the \HI intensity fluctuation power spectrum of NGC~6946 using The HI Nearby Galaxy Survey (THINGS) data. The measured power spectrum has a power law indices of $-1.6\pm0.1$ in the range starting from $4$ kpc to $300$ pc. 
 We present the results of our estimation of \HI column density and line of sight turbulent velocity power spectrum using VME in NGC~6946. We use the combination of the VLA B, C and D configuration data from the THINGS survey and new observations taken from the VLA A array configuration in our analysis. The paper is organized as follows; section~\ref{VME} discuss the brief overview of the VME estimator, section~\ref{obs} describes the observation and further data reduction procedure of NGC~6946. The results of the implementation of VME on calibrated data are presented in section~\ref{results} and inferences on the possible driving mechanism behind the turbulence are discussed in section~\ref{discu}. We conclude the paper in section~\ref{conc} where we present our main findings of this work.
%=====================================================================
%=============================== Section 2 ==============================
%=====================================================================
\section{Brief overview on Visibility Moment power spectrum Estimator (VME)}
\label{VME}
Characteristics of a turbulent ISM are imprinted in the two-point correlation functions of its density and velocity fluctuations. These can be probed through observation of \HI 21 cm emission using radio interferometers. During the \HI 21 cm observations, a radio interferometer measures the visibility function $V(\vU,\nu)$ at the baseline $\vU$ in the observing frequency $\nu$. The baseline is defined as the projected distance between antenna pairs in units of observed wavelengths. For a considerably small portion of the sky, these measured visibilities can be approximated as the Fourier transform of the specific intensity $I(\vt,\nu)$ of \HI emission coming from any point $\vt = (\theta_x,\theta_y)$ in the sky \citep{2017isra.book.....T}. Here, $\theta_x$ and $\theta_y$ are the cartesian coordinates in the small patch of the observed sky with the origin at the centre of the field of view of observation. The visibilities are measured only at discrete points sampled by antenna pairs. Since for most of the radio interferometers, the visibilities are not sampled uniformly and regularly in the baseline plane, the observed visibilities provide rather incomplete information about the sky in its Fourier plane. Deconvolution of the sampling pattern is used to generate a specific intensity distribution model of the sky from measured visibilities, however, incomplete baseline coverage of an interferometer makes the
deconvolution not an effective tool for estimates of two-point correlations \citep{2019RAA....19...60D}.  The two-point correlations can be estimated unbiasedly from the observed visibilities using  Visibility Moment Estimator (VME) as follows. 
 
The Visibility Moment Estimator (VME) estimate the power spectrum of the \HI column density and line of sight turbulent velocity of external spiral galaxies using the moments of visibility.   \cite{2009MNRAS.398..887D} estimates the  power spectrum of \HI intensity fluctuations using visibility correlation estimator and later  \citep{2019RAA....19...60D} studies the advantage of using this estimator in obtaining unbiased power spectrum measurement. 
The VME also uses this method of correlating measurement at the visibility domain for estimating the power spectrum. In a typical interferometric observation though it is expected that visibility correlations are affected by antenna beam pattern, from previous demonstrations of visibility  correlation estimator,  it is shown that for a galaxy having angular extent of $\theta_0$, these effects can be mitigated at baselines larger than $1/\theta_0$ \citep{2001JApA...22..293B,2009MNRAS.398..887D}. 
 A detailed description of how this is taken care of in VME and their validity tests can be found in \cite{2016MNRAS.456L.117D} and \cite{2020MNRAS.496.1803N}. We discuss only the basic foundation of these estimators here. The estimators start by taking a zeroth and the first moment of the visibilities with respect to the observing frequencies. 
The $j$th moment of the visibility function is defined as 
%\begin{equation}
$V_j (\vU) = \int \id \nu\, \nu^{j}\ V(\vU, \nu).$
%\end{equation}
These moments give information about the column density $(j=0)$ and line of sight velocity distributions $(j=1)$. The column density distribution has an imprint on the overall large-scale \HI structure of the galaxy as well as the fluctuations due to compressible turbulence. The estimator uses the reconstructed image of the galaxy to estimate the large-scale distribution of \HI column density unbiasedly and mitigate its effect from the visibility moment zero $(j=0)$ in the baseline plane to unbiasedly estimate the column density power spectrum. The visibility moment one (i.e, $j=1$) incorporates the information on column density, the line of sight component of the galaxy rotation velocity and the random turbulent velocity component. The VME uses the estimated column density power spectrum from visibility moment zero and the estimates of the galaxy rotation velocity to find the power spectrum of the line of sight component of turbulent velocity fluctuations. One of the major assumptions of VME is that the galaxy of interest is a thin relatively face-on disk with an inclination angle between $\sim 15^{\circ} - 40^{\circ}$. NGC~6946 has an average inclination of $33^{\circ}$, well within this requirement. %The following section details the observation and preparation of data {\bf for NGC~6946}.

\section{Observation and data reduction}
\label{obs}
NGC~6946 is a SABcd-type bright late-type spiral galaxy with prominent spiral structures \citep{1991rc3..book.....D,2014MNRAS.444L..85F}. Using the luminosity-based estimate by \citet{2004AJ....127.2031K} we adopt a distance of $5.9$ Mpc to the galaxy in our analysis. \HI emission from the NGC~6946 has been observed as part of the THINGS survey \citet{2008AJ....136.2563W}. The rotation curve presented in \cite{2008AJ....136.2648D} shows that the systematic tangential velocity of NGC~6946 is fairly constant for a galactocentric radius greater than $6$ kpc. Moreover, the approaching and receding components of the tangential velocity are also quite similar indicating a flatter disc. The dynamical inclination and position angles are found not to vary considerably making it a good choice for the application of VME. The \HI line profile width at $20$\% of the peak intensity for NGC~6946 is $240$ km s$^{-1}$. Table \ref{tab1} summarizes the various parameters and observational details of NGC~6946: (1) distance to the galaxy, (2) \HI inclination angle ($i$), (3) position angle ($p$), (4) optical radius ($r_{25}$), (5) \HI major and minor axes, (6) star formation rate, (7) the total \HI mass, (8) dynamical mass and column (9) to (13) gives the details about the new \HI observation carried out using VLA telescope. 
The inclination and position angles given are disc averaged values and adopted from \cite{2008AJ....136.2648D}. The optical radii $r_{25}$ is the radius where the B band surface brightness is $25$ magnitude arcsec$^{-2}$ \citep{2008AJ....136.2563W}. The \HI extent is defined as the major and minor axis at the level of the column density of $10^{19}$ atoms cm$^{-2}$ \citep{2013NewA...19...89D}.
The reasonably larger extent of the \HI disc of the galaxy helps us to measure the power spectrum to larger length scales.
\begin{center}
\begin{table}
\setlength{\tabcolsep}{6pt} % Default value: 6pt
\renewcommand{\arraystretch}{1} % Default value: 1
\begin{tabular}{ll}
	\hline
\multicolumn{2}{c}{NGC 6946}                                        \\
	\hline
(1) Distance                   & $5.9$ Mpc                          \\
(2) Inclination angle          & $33^{\circ}$                       \\
(3) Position angle             & $243^{\circ}$                      \\
(4) Optical radius ($r_{25}$) & $9.8$ kpc                          \\
(5) \HI extent  & $35'\times \ 25'$                \\
(6) Star formation rate        & $2.52 \  \rm M_{\odot}\rm yr^{-1}$ \\
(7) \HI mass    & $4.1 \times 10^9\rm M_{\odot}$    \\
(8) Dynamical mass             & $7.3 \times 10^{11}\rm M_{\odot}$ \\
(9) VLA observation cycle   & VLA 18/A, 19/A \\ 
(10) Array Configuration      & A  (3$\sim$170 k$\lambda$)    \\ 
(11)  Bandwidth                    & 4 MHz  (L Band)\\
(12) Spectral resolution	& $\sim 0.4$ km s$^{-1}$ \\
(13) On source time 		& 11 hour 36 min \\
	\hline
\end{tabular}
\caption{Table \ref{tab1} summarizes the various parameters and observational details  of NGC~6946: (1) distance to the galaxy, (2) \HI inclination angle ($i$), (3) position angle ($p$), (4) optical radius ($r_{25}$), (5) \HI major and minor axes, (6) star formation rate, (7) the total \HI mass,  (8) dynamical mass and column (9) to (13) gives the details about the new \HI observation carried out using VLA telescope. 
The inclination and position angles given are disc averaged values and adopted from \citet{2008AJ....136.2648D}. The optical radii $r_{25}$ is the radius where the B band surface brightness is  $25$ magnitude arcsec$^{-2}$ \citep{2008AJ....136.2563W}. The \HI extent is defined as the major and minor axis at the level of the column density of  $10^{19}$ atoms cm$^{-2}$ \citep{2013NewA...19...89D}. 
}
\label{tab1}
\end{table}
\end{center}

As a part of the THINGS survey, NGC~6946 was observed using B, C, and D array configurations of VLA for $10.5$, $1.75$ and $0.25$ hours respectively with a spectral resolution corresponding to $2.6$ km s$^{-1}$. 
Using calibrated visibilities from the THINGS survey, \citet{2013NewA...19...89D} estimated specific intensity fluctuation power spectra of NGC~6946 by directly using visibility correlations. They find a single power law of slope  $-1.6\pm0.1$ in length scale ranging from $300$ pc to $4$ kpc that fits the \HI intensity fluctuation power spectrum. This was considered as an indication of turbulence cascade in the disk. Due to the limited 
sensitivity and spectral resolution across required ranges of length scales, THINGS data itself is not adequate enough for the estimation of the line of sight velocity power spectrum. To probe the column density power spectrum to much lower scales as well as estimate the line of sight velocity power spectra we proposed to observe the \HI emission from this galaxy with the VLA A array configuration. The intention is to add the information from the VLA A array baseline configurations that have a higher spectral resolution with the already existing baseline coverages from the VLA B C and D arrays. We observed \HI emission from the NGC~6946 using the A array configuration having baseline coverage ranging from $3 \ k\lambda$ to $170\ k\lambda$ at L-Band in two observation cycles in 2018 and 2019. A total of $14$ hours of observation was broken down into separate $2$ hours long observation blocks and carried out in a total of seven days during the two cycles. In total, we observed the galaxy for a total of $11.6$ hours in the A array configuration. We use a total bandwidth of $4$ MHz ($\sim 845$ km s$^{-1}$) with $2048$ channels giving a spectral resolution of $~\sim 0.4 $ km s$^{-1}$. The \HI emission from our Galaxy contributes within the part of the bandwidth of observation, giving rise to additional emission from the direction of the galaxy and absorption in the spectra of the used bandpass calibrator source. We implemented in-band frequency switching for bandpass calibration and hence accurately spectral calibrated the data. We also identified the channels with extra emissions from our galaxy and flagged them from the observed data.
The primary calibration and flagging were carried out in each observational block separately using CASA \footnote{Common Astronomy Software Applications}.
 We combined all the observational blocks into a single data using task `concat' in CASA. Final self-calibration and continuum subtraction procedures were done in this combined data. For continuum subtraction, we make a model of the continuum and subtract it in the uv-space from the visibilities.
 We use the task  `ft' and `uvsub'  for the continuum subtraction. The calibrated and continuum-subtracted data is then used for further analysis. We also generate a natural weighted CLEAN image cube from the continuum subtracted visibilities and generate the required moment maps from this image (see \cite{2019RAA....19...60D} for detail). As mentioned earlier, we include the combined B, C and D array data from the THINGS survey in our analysis. We received the primary calibrated visibility data from the THINGS survey from \citet{2008AJ....136.2563W} on request. We performed visibility-based continuum subtraction using a similar procedure described here for THINGS data. The final calibrated and continuum-subtracted data from both observations are further used for power spectrum estimations.

%  Note that the largest baseline measurable in A  array configuration corresponds to an angular scale $----$" which is rather small compared to \HI angular extent of NGC~6946 ($25'\times35'$). We estimate the powre   Hence we use the BCD configuration data of THINGS survey to improve the sensitivty and baseline coverage. The old and new observations data are processed with our power spectrum estimators and later combined the two estimated power spectra. 
%=========================== New Section ==================================
\section{Results}
\label{results}
\label{sec:results}

Implementation of the visibility moment estimator (VME) is discussed in \cite{2020MNRAS.496.1803N}. We follow the same procedure here with the calibrated and continuum subtracted visibilities. We estimate the column density and velocity power spectrum for our new VLA A band observation and the combined B, C and D array observation from the THINGS survey separately. 
To calculate the locally averaged intensity moments, we use a two-dimensional Gaussian kernel having width $7'$ which is roughly the one fourth of the \HI extent in the disc.
% which is defined as $\frac{1}{2\pi\theta_0^2}\exp(-|\vec{\theta}|^2/2\theta_0^2)$  with $\theta_0=3.5'$ 
 Hence estimated locally averaged intensity moments are used to mitigate the large-scale variation of velocity and density from the estimated visibility correlations as a part of the VME procedure \cite{2013MNRAS.436L..49D}. %{\bf  use this reference in introduction somewhere.} 
The errors of the power spectra are estimated by following the methods used in \cite{2011arXiv1102.4419D}.

%The power spectrum of column density as well as line of sight velocity fluctuations are real quantities by definition. As the visibility correlations are done at nearby baselines instead of the same baselines to avoid noise bias \citep{2013MNRAS.436L..49D}, the final annular averaged measured quantities are complex. Since visibility is a real quantity, the imaginary part of the power spectrum is expected to be negligible and hence we consider only points where the real part is $3$ times greater than the imaginary part. 

\begin{figure}
\begin{center}
\includegraphics[width=\linewidth]{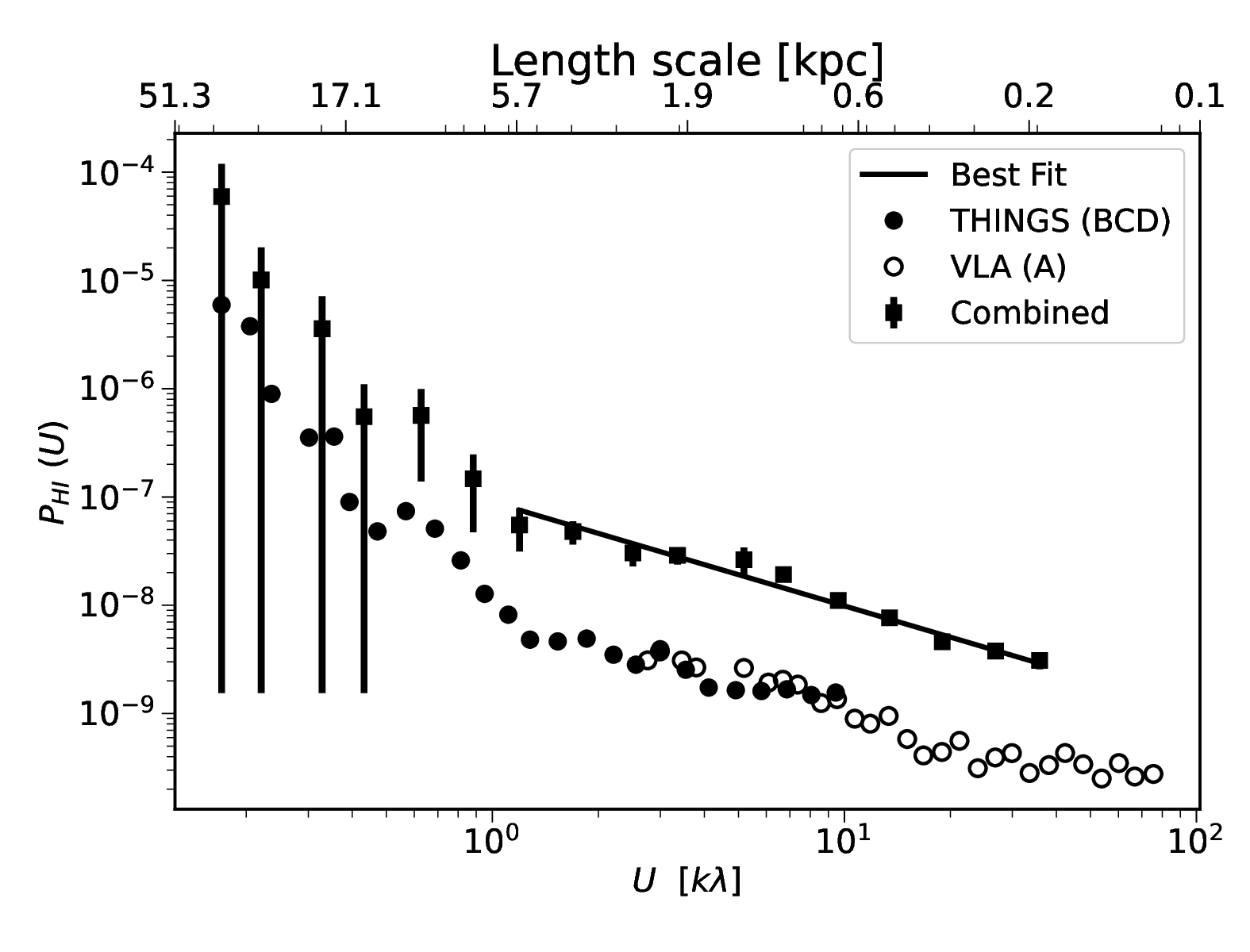}
	\caption{Column density power spectrum of NGC~6946 as a function of baseline $U$. The corresponding length scales are also shown in the top margin. The black solid circles correspond to the measurement of the column density from the THINGS data alone. The black open circles correspond to measurement using only the new VLA observations. These two sets of points are scaled down by a factor of ten for display purposes. The black solid squares with error bars give the power spectra measurements combining the old THINGS and the new observations. A best-fit power law to the combined measurement is shown using black solid lines.}
\label{fig:fig1}
\end{center}
\end{figure}
\begin{table}
\setlength{\tabcolsep}{1pt} % Default value: 6pt
\begin{tabular}{|ccccc|}
\hline
\multicolumn{1}{|c|}{}                      & \multicolumn{1}{c|}{$P_{\hi}(\vec{U})$} & \multicolumn{1}{c|}{$P_{\hi}(\vec{U})$} & \multicolumn{1}{c|}{$P_{\hi}(\vec{U})$} & $P_{v}(\vec{U})$    \\ \hline
\multicolumn{1}{|c|}{Data}                  & \multicolumn{1}{c|}{THINGS}            & \multicolumn{1}{c|}{New Obs}           & \multicolumn{1}{c|}{Combined}          & Combined             \\ \hline
\multicolumn{1}{|c|}{$A (\times 10^{4}) $}  & \multicolumn{1}{c|}{$0.46 \pm 0.08$}     & \multicolumn{1}{c|}{$2.3 \pm 0.5$}    & \multicolumn{1}{c|}{$0.67 \pm 0.09 $}    & $59 \pm  13 ^{*}$ \\ \hline
\multicolumn{1}{|c|}{$\alpha$}              & \multicolumn{1}{c|}{$-0.93 \pm 0.1 $} & \multicolumn{1}{c|}{$-1.09 \pm 0.08$}   & \multicolumn{1}{c|}{$-0.96 \pm 0.05$}  & $-1.81 \pm 0.07$    \\ \hline
\multicolumn{1}{|c|}{reduced $\chi^2$}      & \multicolumn{1}{c|}{0.7}               & \multicolumn{1}{c|}{1.2}               & \multicolumn{1}{c|}{1.1}               & 1.2                \\ \hline
\multicolumn{1}{|c|}{$U_{min} (k\lambda) $} & \multicolumn{1}{c|}{1}              & \multicolumn{1}{c|}{3}                 & \multicolumn{1}{c|}{1}              & 1                \\ \hline
\multicolumn{1}{|c|}{$U_{max} (k\lambda) $} & \multicolumn{1}{c|}{6}                 & \multicolumn{1}{c|}{36}                & \multicolumn{1}{c|}{36}                & 35                  \\ \hline
\multicolumn{1}{|c|}{$R_{min} (kpc)$}       & \multicolumn{1}{c|}{1}                 & \multicolumn{1}{c|}{0.16}               & \multicolumn{1}{c|}{0.16}              & 0.17                \\ \hline
\multicolumn{1}{|c|}{$R_{max} (kpc)  $}     & \multicolumn{1}{c|}{6}                 & \multicolumn{1}{c|}{2}                 & \multicolumn{1}{c|}{6}                 & 6                   \\ \hline

%\multicolumn{1}{|c|}{$A (\times 10^{4}) $}  & \multicolumn{1}{c|}{$0.46 \pm 0.08$}     & \multicolumn{1}{c|}{$2.3 \pm 0.5$}    & \multicolumn{1}{c|}{$0.67 \pm 0.09 $}    & $59.1 \pm  13.5 ^{*}$ \\ \hline
%\multicolumn{1}{|c|}{$\alpha$}              & \multicolumn{1}{c|}{$-0.93 \pm 0.1 $} & \multicolumn{1}{c|}{$-1.09 \pm 0.08$}   & \multicolumn{1}{c|}{$-0.96 \pm 0.05$}  & $-1.81 \pm 0.07$    \\ \hline
%\multicolumn{1}{|c|}{reduced $\chi^2$}      & \multicolumn{1}{c|}{0.7}               & \multicolumn{1}{c|}{1.2}               & \multicolumn{1}{c|}{1.1}               & 1.2                \\ \hline
%\multicolumn{1}{|c|}{$U_{min} (k\lambda) $} & \multicolumn{1}{c|}{1}              & \multicolumn{1}{c|}{3}                 & \multicolumn{1}{c|}{1}              & 1                 \\ \hline
%\multicolumn{1}{|c|}{$U_{max} (k\lambda) $} & \multicolumn{1}{c|}{6}                 & \multicolumn{1}{c|}{36}                & \multicolumn{1}{c|}{36}                & 35                  \\ \hline
%\multicolumn{1}{|c|}{$R_{min} (kpc)$}       & \multicolumn{1}{c|}{1}                 & \multicolumn{1}{c|}{0.15}               & \multicolumn{1}{c|}{0.16}              & 0.17                \\ \hline
%\multicolumn{1}{|c|}{$R_{max} (kpc)  $}     & \multicolumn{1}{c|}{6}                 & \multicolumn{1}{c|}{3}                 & \multicolumn{1}{c|}{5}                 & 6                   \\ \hline

\end{tabular}
\caption{Result of power law fit ($P = A U^{\alpha}$) to the column density ($P_{\hi}$) and velocity ($ P_v$) power spectra of NGC~6946. The power law amplitude ($A$) at one steradian, best fit slope $\alpha$, $1\sigma$ errors associated with the fit, the reduced $\chi^2$ values and the range of fit in baselines and corresponding length scale ranges are shown. The density power spectra fit is shown for the THINGS data, the new observation data and the combined power spectra. The amplitudes for the density spectra are scaled by a factor of $10^4$ and have a unit of steradian. $^*$The amplitude of the velocity spectra is not scaled and is given in (km sec$^{-1}$) $^2$ steradian. }

\label{tab:tab1}
\end{table}

 Using central 400 channels that correspond to $165$ km s$^{-1}$ of the processed visibility data, we estimate the \HI column density and line of sight velocity power spectrum. Note that we only consider the central part of line emission where the signal to noise ratio is much higher. The estimated \HI column density power spectrum of different observations is given in figure~\ref{fig:fig1}. 
Empty circles show the measurements of the column density power spectrum $P_{\hi}$ as a function of baselines using new VLA A array observations and filled circles shows that of THINGS observations. For the THINGS observation also, we use the same part of emission for power spectrum generation.
Note that the power spectrum values shown for new VLA and THINGS observations are scaled down by a factor of $10$ from their original values for presentation purposes. A power law of $P_{\hi} = A U^{\alpha}$ is fitted to these power spectra and fitting parameters are given in table~\ref{tab:tab1}. The column density power spectra estimated from new observation is following a power law from $2$ kpc to $160$ pc with a slope of $-1.09 \pm 0.08$ and that of THINGS observations has a slope 
$-0.93 \pm 0.01$ from $6$ kpc to $1$ kpc. Note that with the new A array observations, we are able to measure the power spectrum up to much smaller length scales. We merged the estimated power spectra over the entire baseline range and divided them into an equal number of bins on a logarithmic scale. The final combined power spectrum is estimated by taking the average of power spectrum estimates in each logarithmic bin and the respective errors in the bins are calculated by taking the root mean square of the individual error of the estimates in each bin.
Black squares with error bars show the combined power spectrum and the black solid line represents the best-fit power law to the same.  
Clearly, the combined power spectra also follow a power law in a baseline range of $1 - 36$ k$\lambda$, which is equivalent to a length scale range from $6$ kpc to $160$ pc in the galactic disc of  NGC~6946. The best fit value of $A$ and $\alpha$ to the combined power spectra are $(6.7 \pm 0.9)\times 10^{-5}$ steradian and $-0.96\pm0.05$ . The amplitudes mentioned here are at the baseline of one wavelength. Note that the combined power spectrum (black squares) shows a clear deviation from the power law at smaller baseline of $< 1$ k$\lambda$. This is because, at these smaller baselines, the effect from the large-scale features of the disc dominates. The slope of the column density power spectrum of NGC~6946 measured in our analysis matches only at $3\sigma$ of uncertainty level with the slope of the intensity fluctuation power spectrum reported by \citet{2013NewA...19...89D}. Note that, the visibility moment estimator of the column density power spectrum suppresses the effect of the window function at lower baselines better than the legacy visibility correlation estimator used earlier. Hence, the slope reported here can be considered more accurate.

\begin{figure}
\begin{center}
\includegraphics[width=\linewidth]{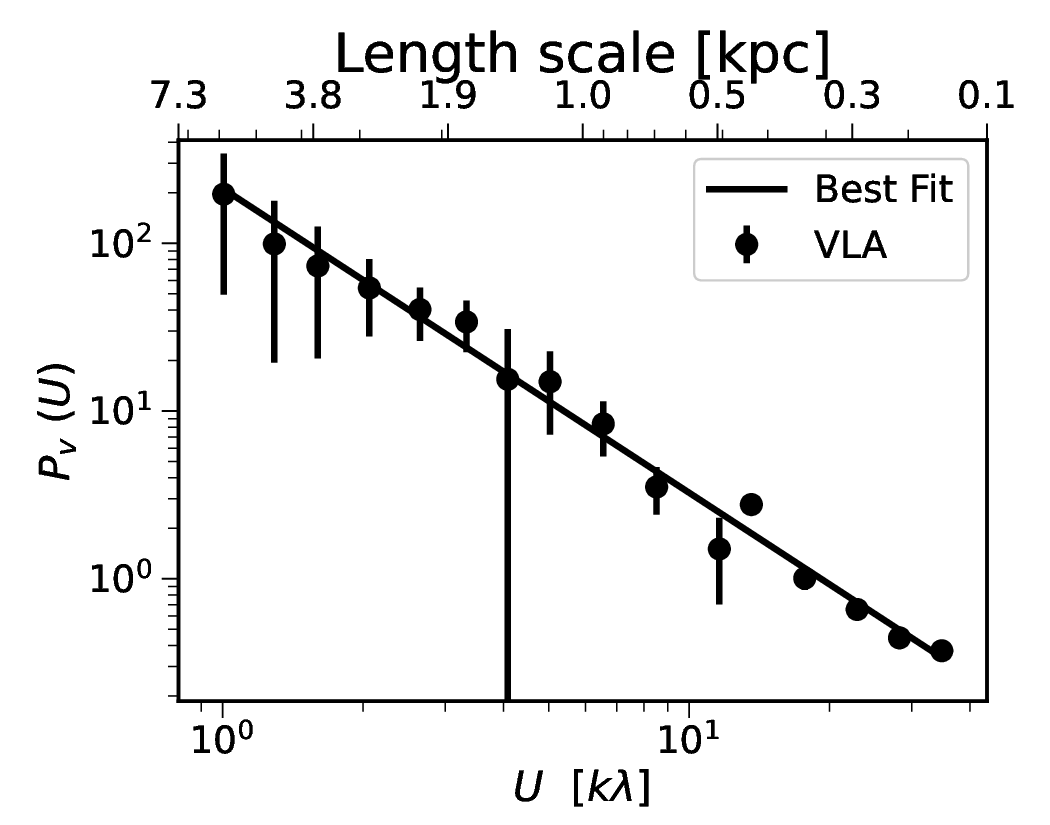}
	%\subcaption{NGC~6946}
\caption{The turbulent velocity power spectrum of NGC~6946 as a function of baseline. The corresponding length scales are also shown in the top margin. The black solid circles with error bars correspond to the combined measurement. A best-fit power law to the combined measurement is shown using black solid lines.}
\label{fig:fig2}
\end{center}
\end{figure}
Similarly, we estimated the line of sight velocity power spectra of THINGS data and new VLA A array observation and combined them to get the final line of sight velocity power spectrum. The power spectrum estimates along with the error bars are shown with black circles in Figure~\ref{fig:fig2}.
The black solid line represents the best-fit power law to the estimated power spectrum values. The velocity spectrum is obeying a power law of amplitude $59\pm13$ (km s$^{-1}$)$^2$ steradian and slope $-1.81\pm0.07$ in the baseline range starting from $1 $ k$\lambda$ to $35 $ k$\lambda$. This observationally confirms the presence of turbulence cascade in the galactic disc of NGC~6946 in scales ranging from $6$ kpc to $170$ pc. 

%=========================== New Section ==================================
\section{Discussion}
\label{discu}
Our measurement of column density and line of sight turbulent velocity power spectrum of \HI in spiral galaxy NGC~6946 indicates the presence of a large-scale energy cascade in the disc. In the power spectra of dust emission in Large Magenallic Cloud (LMC) and nearby galaxy M33 and that of \HI emission of NGC~1058, a break in the spectra has been observed at the scales nearer to the scale height  of the galaxy \citep{10.1111/j.1745-3933.2009.00684.x, 2010ApJ...718L...1B,2012A&A...539A..67C}. 
 \cite{2009MNRAS.398..887D} have demonstrated using numerical simulation that a 2D-3D transition gives rise to a change in the slope of the power spectrum. They find that the relation between the angular scale in which the change in slope happens and the scale height of the disk depends on the disk profile. \cite{2021MNRAS.505.1972K} also reported that the scale at which the power law index changes is not the scale height of the disk but related to it. The scale at which the break is observed is larger than the scale height of the disc (see  \cite{2009MNRAS.398..887D} ). This broken power law is interpreted as a transition from large-scale two-dimensional structures to small- scale three-dimensional structures at the scale height of the disc. \cite{2020MNRAS.492.2663K} show that the change in a slope is also observed at an angular scale related to the PSF of the telescope. This effect is particular to the power spectrum estimation from images due to the existence  of pixels smaller than the telescope resolution \citep{2019RAA....19...60D}. Interestingly, in our measurement, we do not see any transition or break in the power law in both column density and velocity power spectra in the entire range where the cascading is happening. Hence the power law power spectra of both density and velocity can be interpreted as the two-dimensional structure that is generated by turbulence in the disc.

\begin{table}
\setlength{\tabcolsep}{4pt} % Default value: 6pt
\renewcommand{\arraystretch}{1} % Default value: 1
\begin{tabular}{|lccc|}
\hline
%& & & \\
Type of & & & \\
Turbulence   & $P_{\hi}(\vec{U})$  & $P_{v}(\vec{U})$  & References  \\  
%& & & \\
\hline
%& & & \\
Kolmogorov  		&  $-5/3$*                   & $-5/3$            & \citet{1941DoSSR..30..301K}\\ 
Burger's    		& -                   & $-2$              & \citet{BURGERS1948171}\\
	Weizsacker              & $4\alpha - 2$       & $-5/3 - 4\alpha/3$  & \citet{1996ApJ...458..739F} \\
Solenoidal  		& $-0.78\pm0.06$      & $-1.86\pm0.05$    & \citet{2009ApJ...692..364F}\\ 
Compressive 		& $-1.44\pm0.23$      & $-1.94\pm0.05$    & \citet{2009ApJ...692..364F}\\ 
NGC~5236		& $-1.23\pm0.06$      & $-1.91\pm0.08$    &  \citet{2020MNRAS.496.1803N}\\
NGC~6946		& $-0.96\pm0.05$      & $-1.81\pm0.07$    &  This work\\ \hline
\end{tabular}
\caption{Power spectral slope for density and velocity for different theoretical models of turbulence, simulations and observation. %All the spectral slopes are given for what is expected for turbulence in the thin disk as the case for this observational result. 
Here $\alpha$ is the compressibility factor. *Note that though Kolmogorov universal scaling is for incompressible flows, isothermal compressible hydrodynamic simulations for sub/transonic turbulence suggest that the density field also has a similar slope \citep{2005ApJ...630L..45K}.}
\label{tab:comp}
\end{table}

The slope of the power law power spectra of column density and line of sight turbulent velocity is the quantifier that tells us about the nature of the turbulence. For instance, in incompressible fluid turbulence, Kolmogorov theory predicts stated that the slope of the velocity power spectra is $-\frac{5}{3}$ \citep{1941DoSSR..30..301K}. 
In the case of compressible fluid, though a well-established theory is not yet introduced, several analytically computational models are proposed \citep{BURGERS1948171,1951ApJ...114..165V,1996ApJ...458..739F}. The expected slope of the power spectra proposed in these models are given in table~\ref{tab:comp}.  
 \citet{2009ApJ...692..364F} find the slope of column density and velocity power spectra of compressible fluid turbulence driven by different types of forcing using numerical simulations. The slope of \HI column density and turbulent velocity power spectra of NGC~5236 indicate the driving force that generates turbulence therein is compressive in nature \citep{2020MNRAS.496.1803N}. %We compare our slope of estimation with the above-mentioned results in table~\ref{tab:comp}. 
For the galaxy NGC~6946, the density power spectrum have a slope of $-0.96$. Using the scaling relation between density and velocity power spectrum slope for compressible fluid turbulence by \cite{1996ApJ...458..739F}, the velocity power spectrum slope for NGC~6946 is expected to be $-2.01$. This is consistent with our measured slope of $-1.80\pm0.13$ within two sigma uncertainties. 
The slope of the velocity fluctuation power spectra for two different forcing mechanisms, solenoidal and compressive, is rather similar in \citet{2009ApJ...692..364F} analysis and it is difficult to use the measured slope of the velocity fluctuation power spectrum only to comment on the nature of turbulence forcing. However, the density fluctuation assumes different power law slopes for different forcing mechanisms, making the slope of the density fluctuation power spectrum an indicator of the forcing mechanism.
The column density slope of $-0.96$ is intermediate to the slopes found for compressive and solenoidal forcing by \citet{2009ApJ...692..364F}. We believe that the turbulence-driving mechanism for NGC~6946 comes from a combination of compressive and solenoidal forcing. The following subsection discusses the various possible drivers that can attribute the measured turbulence.

%This believe is supported by the fact that NGC~6946 is  
%found to have well designed magnetic spiral arms located between optical arms \citep{1996ARA&A..34..155B}.   

\subsection{Nature of turbulence in NGC~6946 }

\label{subsec:sol}
%\begin{figure}
%\begin{center}
%\includegraphics[width=\linewidth]{Fig3.eps}
%	\caption{Median velocity dispersion is plotted with galactocentric radius (circles) NGC~6946. The vertical line corresponds to the respective $r_{25}$ of the galaxy. The horizontal line corresponds to the turbulence velocity dispersion at $6$ kpc.}
%\label{fig:vdisp}
%\end{center}
%\end{figure}
Comparing the value of slopes of our power spectra estimations with the various theoretical and simulation models of turbulence in the literature, we find that the input energy to the turbulence in NGC~6946 likely comes partly from solenoidal and partly from compressive forcing. The compressive forcing can arise from the self-gravity of the gas on large scales like in the case of NGC~5236 found by \cite{2020MNRAS.496.1803N}. The known sources for solenoidal forcing in the galactic disks are the differential rotation which generates shear and additional influence of magnetic field in the disk \citep{1999ApJ...511..660S,2010A&A...512A..81F}. 
Large-scale galactic magnetic fields are observed in several external spiral galaxies using radio synchrotron emission \citep{2008ApJ...677L..17C,2009RMxAC..36...25K,2013lsmf.book..215B}. The typical field strength observed in spiral arms is $1-10\mu G$ \citep{Sofue1986}. Inter-arm magnetic fields, known as magnetic arms, are also observed. Such magnetic arms are found to be present in NGC~6946 \citep{1996ARA&A..34..155B}. \cite{2007A&A...470..539B} reports multiple strong magnetic spiral arms (having field strength of $\sim 20\mu G$) in NGC~6946 using radio polarization observation. On their investigation of magnetic properties in this galaxy, the measured total magnetic energy density is found to be of a similar order in the inner part of the galactic disc ($<2$ kpc) compared to the outer part (up to $10$ kpc). The ordered magnetic fields are found to be present in NGC~6946 up to a radius of $\sim12$ kpc and regular magnetic fields are detected up to $\sim15$ kpc in \cite{2007A&A...470..539B} observations. These detected ordered fields are aligned along with the spiral structures which indicates the possible interaction of the gas in the arms with a magnetic field.
Applying a nonlinear-turbulent dynamo model to the NGC~6946, through numerical studies \cite{1999A&A...350..423R} shows that observed magnetic structures in NGC~6946 can be well reproduced. These results suggest the observed magnetic structures are the result of turbulent dynamics in the disc. 
Hence we are here to conclude that the magnetic field of NGC~6946 may have a major role to drive its large-scale turbulence.

The power law behaviour in the power spectra shows an energy cascade in an inertial range starting from $6$ kpc to $170$ pc.
For the measured power law slope of $-0.96$ %and amplitude of $6.7\times 10^{-5}$ steradian
for column density fluctuations, the expected slope of the corresponding autocorrelation function is $1.04$. At scales of $6$ kpc, these correspond to a standard deviation in column density fluctuations of $0.01$. For NGC~5236, the inertial range begins at scales of $11$ kpc, where 
the standard deviation in column density fluctuation estimated is $0.011$ at those scales \citep{2020MNRAS.496.1803N}. 
%Interestingly, from their line of sight velocity power spectra estimation, likely scales of injection of the measured turbulence is at $6$ kpc. 
By scaling the estimated standard deviations of density fluctuations compared with the latter, the standard deviation in column density fluctuations at $11$ kpc for NGC~6946 is twice that of NGC~5236.

Though the velocity of gas has three vector components $v_x,v_y,v_z$ in its galactic coordinates, in a real observation what measures is the line of sight velocity $v_{los}$. 
The $v_x,v_y$ are the velocity components along the plane of the disc and $v_z$ is the vertical velocity component. For a typical galaxy, having an inclination angle $i$,
 the line of sight component is given as $v_{los}=-v_y\sin i +v_z\cos i $. Hence the autocorrelation function of the line of sight velocity, $\xi_{los}$, in terms of different velocity components is,
 \begin{eqnarray}
 \label{vlos}
 \nonumber
 \xi_{los}(|\vt'-\vt|) &=&\left< v_{los}(\vt)v_{los}(\vt')\right>  \\
 &=& \xi_{y}\sin^2 i+\xi_z \cos^2 i ,
 \end{eqnarray}
 where the autocorrelation function of the vertical component $v_z$ is $\xi_z(|\vt'-\vt|)=\left< v_z(\vt) v_z(\vt')\right>$ and 
 that of the planar component $v_y$ is $\xi_{y}(|\vt'-\vt|)=\left< v_{y}(\vt) v_{y}(\vt')\right>$. Note that here we assumed that any cross-correlation terms between the on-plane and off the plane motions are negligible. \cite{2010MNRAS.409.1088B} estimate the power spectrum of the different components of velocity fields
from a hydrodynamic galaxy simulation. They show the power spectra of on-plane and vertical components of velocity in length scales from $\sim 1.5 $ kpc to $\sim 10 $ pc, where all three components follow the same power law at smaller scales. At larger scales of the order of kiloparsecs, on-plane components follow the same power law, however, the perpendicular component flattens at larger scales greater than the disc thickness. This is expected in thin disc galaxies, where the large amplitude of the motion of gas happens in the galactic plane. A similar trend of flattening has been found in the power spectrum of the vertical component at similar scales in the numerical simulations studies on gravity-driven turbulence cascade in the disc by \cite{2023arXiv230113221F}. Hence we can infer the turbulent velocity fluctuations in the plane of the disc as follows from eq~\ref{vlos} after assuming that the right term in the equation is negligible.  
In our estimations, we see that the line of sight turbulent velocity fluctuation power spectrum of NGC~6946 has a power law with %amplitude $59$ (km s$^{-1}$)$^2$ at unit steradian and 
a slope of $-1.81$. This corresponds to a slope of $0.19$ to the autocorrelation function of the line of sight velocity $\xi_{los}$ and hence the standard deviation of $27.1$ km s$^{-1}$ in turbulent velocity fluctuations at $6$ kpc scales in the plane of the disc.

%The figure~\ref{fig:vdisp} shows the radial variation of mean \HI velocity dispersion in NGC~6946 as estimated from the moment 2 map. The vertical line marks $R_{25}$ and the horizontal dashed line corresponds to the velocity dispersion in the line of sight component (left-hand side of eq~\ref{vlos}) at 6 kpc scale. %{\it explain in one sentence, velocity dispersion corresponds to off disc component }
%Here also we see that the off the disc component of the velocity dispersion does not have the large velocity fluctuations as measured at scales higher than the thickness of the disc. This again reassures that the $\xi^2_z \cos^2 i$ term is negligible. 

The energy input rate per unit area by the ISM turbulence can be estimated using 
%\begin{equation}
$\epsilon \sim \frac{1}{2} \times (N_{{\hi}_0} m_{\hi} )\times (v^{T})^2 \times v^{T}/L$,
%\end{equation}
where $N_{{\hi}_0}$ is the average column density over the disc, $m_{\hi} $ is the mass of atomic hydrogen and $v^{T}$ is the turbulent velocity fluctuation. The quantity $N_{{\hi}_0} m_{\hi}$ gives the mass of \HI in turbulence per unit area, $L/v^T$ gives the time scale of energy input. We find from the moment 0 map of NGC~6946, the value of $N_{{\hi}_0}$ as $3.7 \times 10^{20}$ atoms cm$^{-2}$. Considering the energy input scale is at $6$ kpc, we find the turbulence energy input $\sim 3 \times 10^{-7}$ ergs cm$^{-2}$ s$^{-1}$. 
%Taking the average energy released in a supernova of $10^{46}$ ergs as kinetic energy, a supernovae rate of one in $100$ years, the average energy input rate by supernovae per unit area in this galaxy is about $3.5\times 10^{-10}$ ergs cm$^{-2}$ s$^{-1}$, quite comparable to the energy inputs in turbulence. 
 For NGC~5236, this corresponds to an energy input of $\sim 1.4 \times 10^{-7}$ ergs cm$^{-2}$ s$^{-1}$ at similar scales. 
%In spite of the different nature of driving mechanisms, the power spectra of the ISM turbulence follow a two-component power law over a large range of length scales.
 For an average rate of a supernova that happens once in 100 years and releases around $10^{51}$ erg \citep{2017MNRAS.472.2975M} of kinetic energy would give an energy input rate per unit area of  $\sim 1\times 10^{-6}$ ergs cm$^{-2}$ s$^{-1}$ in the \HI disc of NGC~6946 after considering an efficiency factor of $\epsilon_{SN} \sim 0.1$ to convert the ejected kinetic energy to turbulent motions \citep{2009AJ....137.4424T,1998ApJ...500...95T}.  Note that the energy input from the supernovae is around six times to the measured turbulent energy at $6$ kpc scales. 
However whether the kinetic energy input from supernovae can reach such large scales coherently is still questionable.

\section{Conclusion }

\label{conc}
It is well established now that the large-scale turbulence cascade is present in the ISM of disc galaxies. Though in small scales it is well clear from both observation and simulations that the stellar feedback is enough to supply energy consistently to the cascade in molecular cloud scales, whether the same is enough to drive the turbulence in the galactic scale where diffuse atomic gas is prominently seen is still in debate. 
  Numerous observational studies on external galaxies support the stellar feedback in sustaining the energy required for large-scale cascades \citep{2020A&A...641A..70B}. \cite{2019ApJ...871...17U} conduct multiwavelength studies of M33 to investigate the origin of ISM turbulence therein, where they did the comparison of calculated turbulent energy with energy input from various sources like supernovae, Magento-rotational instability and gravitational instability. Their results strongly suggest that upto to a galactic radius of 7 kpc, supernovae alone are enough to maintain the observed turbulent energy while the contribution from others is not sufficient. Contradicting to it, \cite{2018MNRAS.479.2505K} shown in the same galaxy that both the energy inputs from supernovae and MRI cannot explain the observed turbulent energy input. 
Various other numerical simulations on turbulent ISM also suggest that stellar feedback is not enough to sustain the turbulence at kiloparsec scales 
\citep{2010MNRAS.409.1088B,2016MNRAS.458.1671K}. Modelling a star-forming ISM that has self-consistent stellar feedback, \cite{2022MNRAS.514.3670C} estimates the power spectrum of coherent density structures where they see that when the turbulence is driven only by stellar feedback, the power spectrum flattens at scales of $60$ pc which is roughly the average length scale of the supernova feedback. \cite{2023arXiv230113221F} introduces numerical simulation on isothermal turbulent cascade in disc galaxy which is purely driven by gravity and estimates power spectra of gas velocity and density. The slope of the velocity power spectra in their estimates obeys Burger's scaling ($\sim -2$ see table~\ref{tab:comp}), similar to our estimates found on both NGC~6946 and NGC~5236.   
Agreement with former results suggests that the compressive component of the driving force to the measured cascade is contributed by gravity. 
One of the possible factors that contribute to the solenoidal part of the driving mechanism is the large-scale ordered magnetic field indirectly coupling with the gas and hence generating an energy cascade.  
 \cite{2017IAUS..322..123F} probed the turbulence in Galactic region G0.253+0.016 and suggest that shear causes the turbulence.  
Hence another possibility here that generates solenoidal turbulence is the galactic shear generated by the differential rotation in the disc. In the lack of sufficient observational results, numerical simulations of the turbulent disc with different types of driving forces should be conducted and compared in order to establish the contribution of various galactic phenomena in creating the energy cascade in the disc.

In this work, we present evidence of the gas turbulence cascade in galactic scales in disc galaxy NGC~6946. We use the visibility moment estimator (VME) to measure the \HI column density and line of sight turbulent velocity power spectrum from VLA observations. Comparing our results from the earlier estimations on NGC~5236, we see that irrespective of the similar inertial range of cascade, the generating mechanism that drives turbulences is different in the two galaxies. To understand the universal picture of the turbulence generation in the interstellar medium of galaxies, measurement of the turbulence cascade at galactic disc on a large number of galaxies and interpreting hence obtained results with various numerical simulations is required. We would like to proceed in this direction in future.

{\it }

%\subsection{Comparison with previous results}
%\label{subsec:comp}
\section*{Acknowledgement}
We thank Nirupam Roy for his helpful suggestions during the observations with VLA.
MN acknowledges the Department of Science and Technology - Innovation in Science Pursuit for Inspired Research (DST-INSPIRE) fellowship for funding this work. PD acknowledges DST-INSPIRE faculty fellowship support for this work.  MN acknowledges the postdoctoral fellowship support from Max-Planck-Gesellschaft Partner Group Grant and support from the Indian Institute of Science, Bangalore. We thank the staff of the NRAO that made these observations possible. The National Radio Astronomy Observatory is a facility of the National Science Foundation operated under a cooperative agreement by Associated Universities, Inc. We thank the anonymous referee for suggestions that have improved the presentation of the paper significantly.
\section*{DATA AVAILABILITY}
The uncalibrated raw visibility data of  VLA A array observation of NGC~6946 are publicly available  on NRAO Data Archive website ( \href{https://data.nrao.edu}{https://data.nrao.edu}).
 The final calibrated visibility data used in this paper  will be shared on reasonable request to the corresponding author.

\bibliographystyle{mnras}
\bibliography{references}

\end{document}